\documentclass[11pt]{article}
\usepackage[intlimits]{amsmath}
\usepackage{lineno}
\usepackage{amsfonts}
\usepackage{amssymb}
\usepackage{fancyhdr}
\usepackage{epsfig}
\usepackage{rotating}
\usepackage{lscape}
\usepackage{graphicx}
\newcommand {\Bsg}    {\ensuremath{\B \rightarrow X_s \gamma}\xspace}
\newcommand {\rusl}   {\ensuremath{R_\mathrm{u/sl}}}

\def\B      {\ensuremath{B}\hbox{ }}

\newcommand {\bclnu}  {\ensuremath{\b \rightarrow \c \ell \bar{\nu}}\xspace}

\newcommand {\mx}     {\ensuremath{M_{X}}}
\newcommand {\Q}      {\ensuremath{q^{2}}}
\newcommand {\mb}      {\ensuremath{m_b}\xspace}

\newcommand {\mX}     {\ensuremath{M_{X}}}
\newcommand {\mmiss}  {\ensuremath{m_{miss}^2}\xspace}

\newcommand {\breco}  {\ensuremath{B_\mathrm{reco}}}
\newcommand {\brecoil}{\ensuremath{B_\mathrm{recoil}}}

\newcommand {\beq}    {\begin{equation}}
\newcommand {\beqa}   {\begin{eqnarray}}
\newcommand {\beqn}   {\begin{eqnarray}}
\newcommand {\eeq}    {\end{equation}}
\newcommand {\eeqa}   {\end{eqnarray}}
\newcommand {\eeqn}   {\end{eqnarray}}
\makeatletter
\def\slash#1{{\mathpalette\c@ncel{#1}}} 
\makeatother

\newcommand{\lonesf} {\ensuremath{\lambda_1^{SF}}}
\newcommand{\lbarsf} {\ensuremath{\bar{\Lambda}^{SF}}}
\newcommand{\mbsf} {\ensuremath{m_b^{SF}}}
\newcommand{\mbk} {\ensuremath{m_b^{kin}}}
\newcommand{\mupisq} {\ensuremath{\mu^2_{\pi}}}
\newcommand{\mupisqsf} {\ensuremath{\mu^{2,SF}_{\pi}}}
\newcommand{\mupisqk} {\ensuremath{\mu^{2,kin}_{\pi}}}

\newcommand{\gevccsq}{\ensuremath{{\mathrm{\,Ge\kern -0.1em V^2\!/}c^4}}\xspace}

\newcommand{\BABARPubYear}     {05}

\newcommand{\BABARConfNumber}  {11}
\newcommand{\SLACPubNumber} {11310}

\input babarsym

\newcommand{\btoulnu}{$b\rightarrow u l \nu$}

\renewcommand{\babar}{\mbox{\ensuremath{{\displaystyle B}\!{\scriptstyle A}
{\displaystyle B}\!{\scriptstyle AR}}}}

\newcommand {\Bxlnu}  {\ensuremath{\Bbar \rightarrow X \ell \bar{\nu}}\xspace}
\newcommand {\Bxulnu} {\ensuremath{\Bbar \rightarrow X_u \ell \bar{\nu}}\xspace}
\newcommand {\Bxclnu} {\ensuremath{\Bbar \rightarrow X_c \ell \bar{\nu}}\xspace}

\newcounter{TODO}

\long\def\inst#1{\par\nobreak\kern 4pt\nobreak
    {\it #1}\par\vskip 10pt plus 3pt minus 3pt}

\begin{document}

{\pagestyle{empty}

\begin{flushright}
\babar-CONF-\BABARPubYear/\BABARConfNumber \\
SLAC-PUB-\SLACPubNumber \\
\end{flushright}

\par\vskip 1.2cm
\begin{center}
\Large \bf Measurement of the Partial Branching Fraction for Inclusive Charmless
Semileptonic \boldmath $B$ Decays and Extraction of \boldmath \Vub
\end{center}
\bigskip

\begin{center}
\large The \babar\ Collaboration\\
\mbox{ }\\
\today
\end{center}
\bigskip \bigskip

\begin{center}
\large \bf Abstract
\end{center}
Charmless semileptonic decays, $\Bbar \rightarrow X_u \ell \bar{\nu}$, are studied in a
sample of 232 million \BB\ decays recorded with the \babar\ detector, in
events where the decay of the second $B$ meson is fully reconstructed.
Inclusive charmless decays are selected in kinematic regions where the
dominant background from semileptonic $B$ decays to charm is reduced by
requirements on the hadronic mass \mx\ and the momentum transfer \Q. 
The partial branching fraction for $\Bbar \rightarrow X_u \ell \bar{\nu}$
decays for $\mx<1.7$~\gevcc and $\Q>8$~\gevccsq is measured to be
$\Delta\BR(\Bxulnu) = (0.87 \pm 0.09_{\rm stat}\pm 0.09_{\rm sys}\pm 0.01_{\rm
th})\times 10^{-3}$.
The CKM matrix element $|V_{ub}|$ is determined by using theoretical
calculations of phase space acceptances. Theoretical uncertainties in this
extrapolation are reduced by using the inclusive
$b \to s \gamma$ photon spectrum and  moments of the $b \to c \ell \bar{\nu}$
lepton energy and hadronic invariant mass.
\vfill
\begin{center}

Submitted to the XXII International Symposium on Lepton-Photon Interactions
at High Energy, 30 June---5 July, Uppsala, Sweden.

\end{center}

\vspace{1.0cm}
\begin{center}
{\em Stanford Linear Accelerator Center, Stanford University, 
Stanford, CA 94309} \\ \vspace{0.1cm}\hrule\vspace{0.1cm}
Work supported in part by Department of Energy contract DE-AC03-76SF00515.
\end{center}

\newpage
} 

\begin{center}
\small

The \babar\ Collaboration,
\bigskip

B.~Aubert,
R.~Barate,
D.~Boutigny,
F.~Couderc,
Y.~Karyotakis,
J.~P.~Lees,
V.~Poireau,
V.~Tisserand,
A.~Zghiche
\inst{Laboratoire de Physique des Particules, F-74941 Annecy-le-Vieux, France }
E.~Grauges
\inst{IFAE, Universitat Autonoma de Barcelona, E-08193 Bellaterra, Barcelona, Spain }
% ...

\end{center}\newpage

\section{Introduction}
The principal physics goal of the \babar\ experiment is to establish
\CP violation in $B$ mesons and to test whether the observed effects
are consistent with the predictions of the Standard Model (SM). 
\CP violating effects result in the SM from an irreducible phase in the 
Cabibbo-Kobayashi-Maskawa (CKM) matrix which describes
the couplings of the charged weak current to quarks. An improved
determination of the magnitude of the matrix element
\Vub, the coupling strength of the $b$ quark to the $u$ quark, 
will contribute critically to tests of the consistency of the 
measured angles of the unitarity triangle of the CKM matrix. 

The precise determination of \Vub\ is very difficult as, due to the large
charmed backgrounds, we are forced to measure partial branching fractions and 
extrapolate them to the full phase space by relying on QCD based theoretical
calculations. These calculations make use of specific assumptions and  
are affected by different uncertainties.
It is therefore important to
make redundant measurements by using  
several experimental techniques, and different theoretical
frameworks. Measurements of the inclusive and exclusive  
charmless semileptonic decays of $B$ mesons are sensitive to different
theoretical calculations, and therefore to different  
sources of systematic uncertainties. In addition, exploiting the available
kinematic variables which  
discriminate between rare charmless semileptonic decays and the much more
abundant decays involving charmed mesons, gives  
different sensitivities to the underlying theoretical calculations and assumptions. 

In inclusive measurements,  three kinematic variables are discussed in
the literature, each having its  
own advantages: the lepton energy ($E_\ell$), the hadronic invariant mass
(\mx), and  the leptonic invariant mass squared (\Q). 
The first measurements were restricted to the high end of the lepton
spectrum where theoretical uncertainties are very large and therefore the 
extrapolation to the full spectrum becomes uncertain.
Event selection based on \mx\ and \Q\ allows us to select larger portions  
of phase space, but the underlying theoretical assumptions need to be
carefully evaluated. 

Theoretical studies indicate that it is possible to reduce the theoretical
error on the extrapolation by taking  
advantage of other kinematic variables or applying simultaneous cuts on \mx\
and \Q\ in inclusive \Bxulnu\footnote{Charge-conjugate modes are implied 
throughout this paper.} decays~\cite{Bauer:2001rc}. In fact, while the
\mx\ distribution has a
large usable fraction of events, of the order of $70\%$,  
but depends on the shape function (SF) describing the Fermi motion of the $b$
quark inside the $B$ meson,  
the \Q\ distribution is less sensitive to non-perturbative effects and less
dependent on the calculation. Unfortunately, only a small  
fraction of events (about $20\%$) is usable with a pure \Q\ selection. 
The study presented in~\cite{Bauer:2001rc} shows that a combined cut on \mx\ and
\Q\ may mitigate the drawbacks of the two  
methods while retaining good statistical and systematic sensitivities. 

\babar\ has already published a determination of \Vub from a measurement of
the inclusive charmless semileptonic  
branching fraction $\BR$(\Bxulnu)~\cite{Aubert:2003zw} based on the study of the recoil
of fully reconstructed $B$ mesons and  
applying a kinematic cut on \mx$<$ 1.55 \gevcc, which resulted in the most
precise determination of this quantity.  
Nevertheless the measurement was dominated (17\% on the branching ratio,
{\it{i.e.}} 8.5\% on \Vub)  
by the theoretical uncertainty on the underlying kinematic variable
distributions, and therefore on the  
extrapolation to the full phase space. This result was obtained on a limited
dataset, corresponding to about 80 fb$^{-1}$ of integrated luminosity. 

Measurements of \Vub through inclusive charmless semileptonic decays 
\Bxulnu using a combination of \mx\ and $q^2$ (``\mx-\Q'' analysis) have 
also been presented by \babar\
at ICHEP 2004~\cite{Aubert:2004bq}. In this case, the extrapolation of the branching fraction 
measurements from a limited region of phase space to the full spectrum was
done following Bauer {\it{et al}}~\cite{Bauer:2001rc} (BLL in the
following).

The calculation for charmless semileptonic decays~\cite{DeFazio:1999sv} implemented  
in the \babar\ Monte Carlo simulation was also used to evaluate theoretical uncertainties.
The preliminary result presented in this paper extends the analysis published
in~\cite{Aubert:2004bq}. This new study is based on the same analysis  
strategy, but on a much larger dataset.

This paper is organized as follows:
Section~\ref{sec:datasample} describes the detector, the data sample and the
Monte Carlo simulation, including  
a description of the theoretical model on which our efficiency calculations are based.
Section~\ref{sec:strategy} 
describes the event reconstruction and selection, while in
Section~\ref{sec:mxq2} the results of the
\mx-\Q\ analysis are presented. 
These results are interpreted by using the calculations
from BLL and Neubert {\it{et al}}~(BLNP)~\cite{Lange:2005yw,Bosch:2004th,Bosch:2004cb}.

The SF parameters \mb\ and \mupisq\ described in Section~\ref{sec:vub}
are an essential input for the determination of kinematic acceptances.
In our previous result~\cite{Aubert:2004bq} we used two set of values for
\mb\ and \mupisq, as determined from the  
analysis of the photon spectrum in \Bsg decays from CLEO and Belle,
respectively. For the present studies, we use the Belle result.  
The SF parameters can also be determined from the analysis of the electron
energy and hadronic mass moments  
in \bclnu decays. Therefore, in Section~\ref{moments} we quote results
based on the \babar\ measurements of  
SF parameters with \bclnu moments.

\section{Data Sample and Simulation}
\label{sec:datasample}
The data used in this paper were recorded with the \babar\ 
detector~\cite{Aubert:2001tu}
at the \pep2\ collider.
The total integrated luminosity of the data set is 
210.7 \invfb\ collected on the \FourS\ resonance.  The corresponding number
of produced \BB\ pairs is 232 million. 
We use Monte Carlo (MC) simulations of the \babar\ detector based on
\geant~\cite{geant} to optimize selection criteria and to
determine signal efficiencies and background distributions.

\subsection{Simulation of \Bxulnu\ decays}
\label{sec:theosys}
Charmless semileptonic \Bxulnu\ decays are simulated as a combination of both 
exclusive three-body decays to narrow resonances, $X_u = \pi,  \eta,  \rho,  \omega,   \eta^\prime$,  
and inclusive decays to non-resonant hadronic final states $X_u$.
The simulation of the inclusive charmless semileptonic $\B$ decays into hadronic 
states with masses larger than $2m_{\pi}$  is based on a prescription 
by De~Fazio and Neubert~\cite{DeFazio:1999sv} (DFN), which calculates the
triple differential decay rate,  $d^3\Gamma\,/\,dq^2\,dE_{\ell}\,ds_H$
($s_H=\mX^2$), up to ${\cal O}(\alpha_{\rm s})$  corrections.
The motion of the $\b$ quark inside the $\B$ meson 
is incorporated in the DFN formalism by convolving  the parton-level triple
differential decay rate with a non-perturbative SF. 
The SF describes the distribution of the momentum $k_+$  of the $\b$ quark inside the $\B$ meson.  
The two free parameters of the SF are 
$\lbarsf$ and $\lonesf$. The first relates the $B$ meson mass, $m_B$, to the
$b$ quark mass, $\mbsf = m_B - \lbarsf$, and
$\mupisq = -\lonesf$ is the average momentum squared of the ${\b}$ quark in
the ${\B}$ meson.
The SF parameterization used in the generator is of the form  
 \begin{equation}
  F(k_+) = N (1-x)^a e^{(1+a)x},
 \label{eq:fermi_motion}
 \end{equation}
where $x = \frac{k_+}{\lbarsf} \le 1 $ and $a = -3
(\lbarsf)^{2}/{\lonesf}-1$. 
The first three moments of the SF must satisfy: $A_0 = 1$, $A_1 = 0$ and $A_2 = -\lonesf/3$.

In the simulation 
the hadron system $X_u$ is produced with a non-resonant
and continuous invariant mass spectrum according to the DFN model.
The fragmentation of the $X_u$ system into final state hadrons 
is performed by JETSET~\cite{Sjostrand:1994yb}.
The exclusive charmless semileptonic decays are simulated using 
the ISGW2  model~\cite{isgw2}.
The resonant and non-resonant components are combined such that the total branching fraction is 
consistent with the measured value~\cite{Aubert:2003zw} and that the integrated spectra agree
with the prediction of Ref.~\cite{DeFazio:1999sv}. 
In the evaluation of the associated uncertainty all branching fractions and theory parameters
are varied within their errors.

\subsection{Simulation of Background Processes}
To estimate the shape of the background distributions we make use of simulations of 
$\epem\to\FourS\to\BB$ with the $B$ mesons
decaying inclusively. 
The most relevant backgrounds are due to  \Bxclnu\ events. 
The simulation of these processes uses a Heavy Quark Effective Theory (HQET)
parametrization of form factors for $\Bb\to 
D^{*}\ell\nub$~\cite{Duboscq:1996mv}, and models  for $\Bb\to
D^{(*)} \pi \ell\nub$~\cite{Goity:1995xn}, and  $\Bb\to D
\ell\nub,D^{**}\ell\nub $~\cite{isgw2}.
We also make use of a JETSET simulation of ``continuum'' $e^+e^-\to q\bar{q}$
($q=u,d,s,c$) events.

\section{Event Selection and Reconstruction}
\label{sec:strategy}
The event selection and reconstruction and the measurements of branching fractions 
follow closely the strategy described in Ref.~\cite{Aubert:2003zw} and represent the 
basis for the \mx-\Q\ analysis presented here.

In this paper we study the recoiled $B$ candidate opposite of 
a fully reconstructed $B$ 
in hadronic decay (\breco), which is a moderately pure 
sample of $B$ mesons with known flavor and four-momentum.
We select \breco\ decays of the type  $B \rightarrow \Db  Y$,
where $D$ refers to a charm  meson, and $Y$ represents a collection of
hadrons with  a total  charge of $\pm  1$, composed  of $n_1\pi^{\pm}+
n_2K^{\pm}+ n_3\KS +  n_4\piz$, where $n_1 + n_2 < 6$,  $n_3 < 3$, and
$n_4 <  3$.  Using $D^-$  and $D^{*-}$ ($\Dzb$ and  $\Dstarzb$) as
seeds for  $B^0$ ($B^+$) decays,  we reconstruct about  1000 different
decay chains. Overall, we correctly reconstruct one $B$ candidate in 0.3\%  (0.5\%) of the 
\BzBzb\ (\BpBm) events.
The kinematic consistency of a \breco\ candidate with a $B$ meson
decay is checked using two variables, the beam-energy-substituted mass
$\mes = \sqrt{s/4 - \vec{p}^{\,2}_B}$
 and the energy difference, 
$\Delta E  = E_B  - \sqrt{s}/2$. Here  $\sqrt{s}$ refers to  the total
energy in the \FourS center of  mass frame, and $\vec{p}_B$ and $E_B$ denote
the momentum and energy of the \breco\ candidate in the same frame. 
For signal events the \mes\ distribution peaks at the $B$ meson mass, while
$\Delta E $ is consistent with zero. We  require $\Delta E  = 0$ within  approximately three
standard deviations, as determined from each decay channel.

A semileptonic decay of the other $B$ meson (\brecoil) is identified by the
presence of a charged lepton with momentum in the \brecoil\ rest frame ($p^*$) above 1\gevc\ . 
In addition, the detection of missing
energy and momentum in the event is taken as evidence for the presence
of a neutrino. 
The hadronic system $X$ is reconstructed from charged
tracks  and energy depositions
in the calorimeter that are not associated with
the \breco\ candidate or the identified lepton. Care is taken to eliminate 
fake charged tracks, as well as low-energy beam-generated photons and energy
depositions 
in the calorimeter from charged and
neutral hadrons due to beam backgrounds. The
neutrino four-momentum $p_{\nu}$ is estimated from the
missing momentum four-vector $p_{miss} = p_{\Upsilon(4S)}-p_{\breco} -p_X-p_\ell$, 
where all momenta are measured in the laboratory frame, and
$p_{\Upsilon(4S)}$ refers to the \FourS\ meson. 

Undetected particles and measurement uncertainties affect the determination 
of the four-momenta of the $X$ system and neutrino, 
and lead to a leakage of \Bxclnu\ background from the high \mx\
into the low \mx\ region ($\mx < 1.86$~\gevcc).

In the sample of reconstructed $B$ decays (\breco) two backgrounds need
to be considered: the combinatorial background from \BB\ and continuum events, 
due to random association of tracks and neutral clusters, which does not peak
in \mes, and the \BB\ background whose \mes\ distribution has the same shape as the signal.
After applying all selection criteria, the remaining combinatorial
background is subtracted by performing an unbinned likelihood fit
to the \mes\ distribution. In this fit, the combinatorial
background originating from $e^+e^-\to q\bar{q}$ ($q=u,d,s,c$) continuum
and \BB\ events is described by an empirical threshold function~\cite{Albrecht:1993pu}, and 
the signal is described by a modified Gaussian~\cite{cry} peaked at the $B$ meson mass. 
In addition, to further reduce the effects of the combinatorial background, 
only events with $\mes >$5.27~\gevcc are considered.

To reject the background in the sample of semileptonic decays we require exactly one
charged lepton with $p^*>$1\gevc, 
a total event charge of zero, and a missing mass consistent with zero
($\mmiss < 0.5$\gevccsq).  These criteria partly suppress the dominant
\Bxclnu\ decays, many of which contain an additional neutrino or
an undetected $\KL$ meson. 

In order to reject the peaking background coming from $\Dstar$ we use
a dedicated method, the partial $\Dstar$ reconstruction.
We explicitly veto
the $\Bzb\to\Dstarp\ell^-\overline{\nu}$ background by searching candidates 
for such a decay with a partial reconstruction technique, that is 
only identifying
the $\pi^+_s$ from the $\Dstarp\to \Dz\pi_s^+$ decay and the
lepton: since the momentum of the $\pi^+_s$ is almost collinear with the
\Dstarp\ momentum in the laboratory frame, we can approximate the
energy of the \Dstarp\ as $E_{\Dstarp}
\simeq m_{\Dstarp} \cdot E_{\pi_s} /145 \mevcc$ and estimate the neutrino mass as
  $  m_{\nu}^2  =  (p_B   -  p_{\Dstarp}  -  p_{\ell})^2 $. Events with $  m_{\nu}^2 >- 3$~\gevccsq\
are likely to be background events and are rejected.
Finally, we veto events with charged or neutral kaons in the
recoiling \Bb\ to reduce the peaking background from \Bxclnu\ decays. Charged
kaons are identified \cite{Aubert:2001tu}
with an efficiency varying between 60\% at the highest (4~\gevc) 
and almost 100\% at the
lowest momenta. The pion misidentification rate is about 2\%. The
$\KS\to\pi^+\pi^-$ decays are reconstructed with an efficiency of
$80\%$ from pairs of oppositely charged tracks with an invariant mass
between 486 and 510~\mevcc.

\section{Measurement of Charmless Semileptonic Branching Ratios}

\begin{figure}[t!]
 \begin{centering}
\epsfig{file=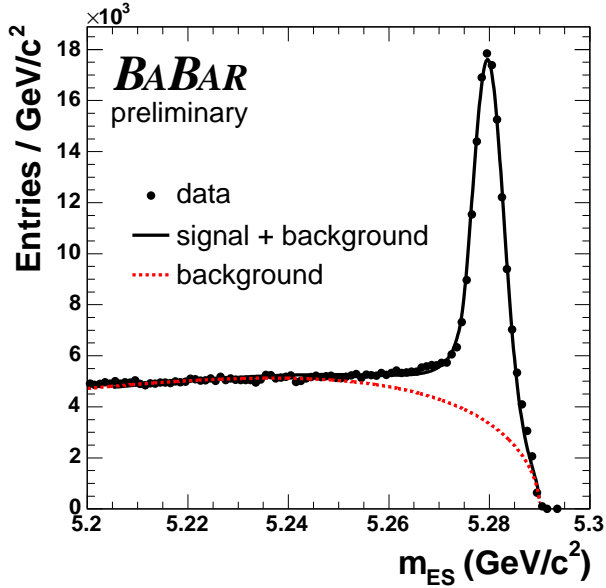,width=0.5\linewidth}
\caption{ Fit to the $\mes$ distribution for the sample with high momentum lepton. 
\label{fig:mxhadfit}}
 \end{centering}
\end{figure}

To reduce the systematic uncertainties in the 
derivation of branching fractions, the observed
number  of  signal events, corrected for peaking background and efficiency, is
normalized to the total number of semileptonic decays \Bxlnu\
in the recoil of the \breco\ candidates. 
The number of observed \breco\ events which contain a charged
lepton  with $p^*>$ 1\gevc is denoted as
$N_{sl}^{meas}$. It can be related to the true number of semileptonic
decays, $N_{sl}^{true}$ and the remaining peaking
background $BG_{sl}$, estimated with Monte Carlo simulation, by
$N_{sl}^{true} = (N_{sl}^{meas} - BG_{sl})/\epsilon_l^{sl} \epsilon_t^{sl}=
N_{sl}/\epsilon_l^{sl} \epsilon_t^{sl}$ .
Here $\epsilon_l^{sl}$ refers to the efficiency for selecting a lepton
from a semileptonic B decay with a momentum above $p_{cut}$ in an
event with a reconstructed $B$ with efficiency $\epsilon_t^{sl}$. Figure \ref{fig:mxhadfit} shows the 
result of the \mes\ fit used to determine $N_{sl}^{meas}$. 

If we denote as $N_u^{meas}$ the number of  events fitted in the
sample  after all requirements, and  with $BG_u$ the peaking background
coming from semileptonic decays other than the signal, the true number of signal events $N_u^{true}$
is related to them by
\begin{equation}
N_u^{meas} - BG_u = \epsilon_{sel}^u  \epsilon_l^u \epsilon_t^u N_u^{true} ,
\end{equation}
\noindent
where the signal efficiency $\epsilon_{sel}^u $ accounts for all selection
criteria applied on the sample after the requirement of 
a high momentum lepton.  

To measure $BG_u$ in the inclusive studies, the peaking background ($BG_u$)
is estimated 
by performing a $\chi^2$ fit on 
the \mx-\Q\ distributions, resulting from \mes\ fits in individual \mx-\Q\ bins, 
with the shape of the background estimated from Monte Carlo simulation,
and its normalization free to vary. 

The ratio between the partial branching fractions for the signal and $\Bxlnu$ decays is 
\begin{equation}
  R_{u/sl} = 
  \frac{\Delta\BR({\rm { signal)}}}{\BR(\Bxlnu)}=
  \frac{N_u^{true}}{N_{sl}^{true}} = 
  \frac{(N_u^{meas}- BG_u)/(\epsilon_{sel}^u)}{(N_{sl}^{meas}-BG_{sl})} 
  \times \frac{\epsilon_l^{sl} \epsilon_t^{sl} } {\epsilon_l^u \epsilon_t^u }.
  \label{eq:ratioBR}
\end{equation}

\noindent
The efficiency ratio 
$\frac{\epsilon_l^u \epsilon_t^u}{\epsilon_l^{sl} \epsilon_t^{sl}}$ 
is expected to be close to, but not equal to
unity. Due to the difference in multiplicity and the different
lepton momentum spectra, we expect the tag efficiency $\epsilon_t$
and lepton efficiency $\epsilon_l$ to be slightly different for the
two classes of events, the largest effect coming from $\epsilon_l$. 
The ratio was measured to be $1.204\pm0.033$.
The signal branching fraction is then obtained from $R_{u/sl}$ using  
the total semileptonic branching fraction of $\BR(\Bxlnu)=(10.83\pm 0.19)\%$,
which is the sum of the charm semileptonic branching ratio  
$\BR(\Bxclnu)= (10.61 \pm 0.16 (\rm exp) \pm 0.06 (\rm th))\%$~\cite{Aubert:2004aw}
and the charmless semileptonic branching ratio $\BR(\Bxulnu)= (0.22 \pm 0.04 (\rm exp.) \pm 0.04 (\rm th))\%$~\cite{Aubert:2003zw},
both measured in \babar.

\subsection{Measurement of the Partial Branching Fraction}
\label{sec:mxq2}

Measurements done using only a \mx\ kinematic cut to reject the \bclnu 
background are limited by the dependence on the  
SF. This can be overcome by selecting a phase space region where the 
SF effects are small, namely the region at large
\Q\ values~\cite{Bauer:2001yb}. 
In this way we find a trade-off between the statistical 
and theoretical uncertainties by loosening the \mx\ cut and applying a
cut on \Q.
Moreover, since most of the theoretical uncertainties are due to the 
extrapolation from a selected kinematic region to the full phase space,  
measurements of partial branching fractions in different regions of phase
space and their extrapolation to the  
full phase space can serve as tests of the theoretical calculations and models. 

\begin{figure}
  \centerline{\epsfig{figure=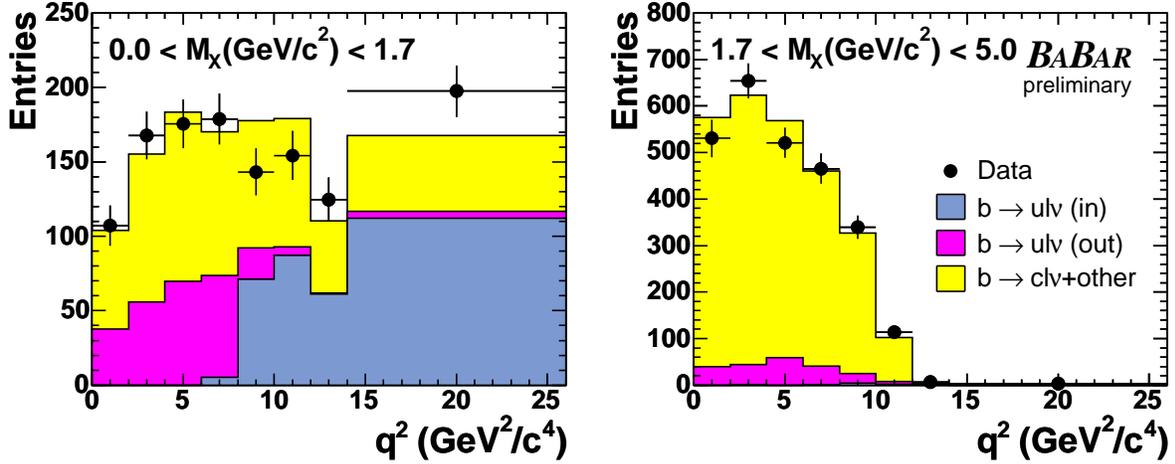,width=16.cm}}
  \caption{Distributions of \Q\ in two intervals of \mx. Points are
    data, the blue (medium shaded), magenta (dark shaded) and yellow (light shaded) histograms  
    represent  the fitted contributions from \btoulnu\ events inside
    true  \mx$<1.7\gevcc$, \Q$>8\gevccsq$, \btoulnu\
    events outside these requirements, and background events, respectively.}
  \label{fig:mxq2fit}
\end{figure}

In order to extract the partial charmless semileptonic branching ratio,
$\Delta \BR(\Bxulnu)$, in a given region of the \mx-\Q\ plane, we define as signal
the events with true values of the kinematic variables in the chosen
region, treating as background those that migrate from outside this region
because of the resolution.
This means that in applying Eq.~\ref{eq:ratioBR} we include
the \btoulnu\ events outside the signal region in $BG_u$ and the 
quoted efficiencies refer only to events 
generated in the chosen (\mx-\Q) region. 
These efficiencies are computed on simulation based on 
the DFN model. However, the associated theoretical uncertainty on the final
result is small compared to the 
extrapolation error to the full phase space. 
We divide the events into 32 non-equidistant two-dimensional bins of \mx\ and \Q\ (4 bins in
\mx\ and 8 in \Q), 
we fit the \mes\ distribution to extract the yield in each bin, and 
we perform a two-dimensional binned fit of the entire \mx-\Q\ distribution 
in order to extract the signal and background components. The result of the fit 
is shown in Fig.~\ref{fig:mxq2fit}. 
We measure, out of $103590 \pm 474$ background-subtracted
semileptonic events $(N_{sl}^{meas}-BG_{sl})$, $317 \pm 34$ signal events $(N_u^{meas}- BG_u)$, above a background of $270
\pm 5$ events $(BG_u)$. This, with  $\epsilon_{sel}^u = 0.319\pm0.006$, corresponds to a
partial branching fraction in the signal region $\Q>8\gevccsq$, $\mx<1.7\gevcc$
of: 
\begin{eqnarray}
  \begin{array}{rll}
    \lefteqn{\rusl(\Bxulnu,\mx<1.7 \gevcc, \Q>8  \gevccsq)  = } \\ 
    & & = (0.80 \pm 0.09_\mathrm{stat} \pm 0.09_\mathrm{syst} \pm 0.01_\mathrm{th}) \times 10^{-2}, 
    \label{mxq2rusl}
  \end{array}
\end{eqnarray}
where the errors are due to statistics, experimental systematics and
theoretical systematics, respectively. This gives the following value for the partial branching fraction: 
\begin{eqnarray}
  \begin{array}{rll}
    \lefteqn{\Delta \BR(\Bxulnu,\mx<1.7 \gevcc, \Q>8  \gevccsq)  =}  \\ 
    & & = (0.87 \pm 0.09_\mathrm{stat} \pm 0.09_\mathrm{syst} \pm 0.01_\mathrm{th}) \times 10^{-3}.
    \label{mxq2res}
  \end{array}
\end{eqnarray}

\subsection{Systematic Uncertainties}
\label{sec:syst}
A breakdown of the systematic uncertainties is presented in
Table~\ref{sysall}.

Uncertainties related to the reconstruction of charged tracks
are  determined  by  removing randomly a fraction of tracks
corresponding to the uncertainty in the track finding efficiency
($1.4\%$ per track). 

For photons, 
we correct for differences between data and MC in 
energy resolution, energy scale, and EMC crystal edge effects and 
assign the systematic uncertainty by repeating the analysis without 
applying the corrections. 
For single photon reconstruction no efficiency correction is applied,
but a systematic uncertainty of 1.8\% per photon is assigned.

We  estimate the systematic  error due  
to particle  identification (PID) by
varying the electron and kaon identification efficiencies  by $\pm 2\%$
and the muon identification efficiency by $\pm3\%$. The
misidentification probabilities are varied  by $15\%$ for all particles. 
Effects due to $K_L$ interactions have been estimated by removing all EMC deposits due to $K_L$
when reconstructing \mx.

The uncertainty of the \breco\ combinatorial background subtraction
is estimated by changing the signal shape function to a Gaussian function instead 
of the empirical function of Ref.~\cite{cry}. Furthermore, the parameters of 
the empirical function, which are kept fixed in the \mes\ fits,
are varied within their uncertainties. 

The size of the Monte Carlo sample limits the accuracy on the
determination of the ratio 
$\frac{\epsilon_l^u \epsilon_t^u}{\epsilon_l^{sl} \epsilon_t^{sl}}$ to 3\%.

The impact of the charm semileptonic branching fractions has been estimated by varying each of
the exclusive  branching fractions within  one standard deviation of the
current world average~\cite{PDG04}. 
Similarly, the branching fractions
of charm mesons for inclusive kaon production have been varied
to estimate uncertainties in the kaon veto. 

To study the mixture of resonant decays among the 
charmless modes we also varied the number of 
charmless exclusive semileptonic decays by $30\%$ for $\Bbar \ra \pi \ell \bar{\nu}$
and $\Bbar \ra \rho \ell \bar{\nu}$, by $40\%$ for $\Bbar \ra \omega \ell \bar{\nu}$ and by 
$100\%$ for the remaining exclusive charmless semileptonic B decays.
Using only the non-resonant model for the signal gives an estimate of the effects due to
uncertainties in the hadronization model.
Signal events where a gluon splits in a $s{\bar{s}}$ pair are varied by 30\% in non-resonant
events in order to obtain the associated systematic uncertainty.

The uncertainties related to the knowledge of the SF are calculated by
changing the functional form and varying the SF parameters as described in Section~\ref{sec:vub}.

\begin{table}
\begin{center}
\caption{Systematic uncertainties in percent for the partial 
fraction $\Delta\BR(\Bxulnu)$.}
\vspace{0.1in}
\begin{tabular}{|l|c|} 
\hline
Source & \raisebox{3.8mm}{\rule{0pt}{1ex}}\raisebox{-4mm}{\rule{0pt}{1ex}}$\frac{\sigma(\Delta\BR(\Bxulnu))}{\Delta\BR(\Bxulnu)}$ \\  \hline
Statistical error         &   \multicolumn{1}{|r@{\hspace{5mm}}|}{10.7}     \\ \hline
Monte-Carlo statistics    &    \multicolumn{1}{|r@{\hspace{5mm}}|}{4.0}     \\ \hline
\multicolumn{2}{|c|}{Detector-related:} \\ \hline
Tracking efficiency       &    2.2     \\
Neutral efficiency        &    0.1     \\
Neutral corrections       &    0.6     \\
$K_L$                     &    2.0     \\
PID efficiency \& misidentification  &    2.5     \\ \hline
Detector uncertainties    &    \multicolumn{1}{|r@{\hspace{5mm}}|}{3.9} \\ \hline
\multicolumn{2}{|c|}{\breco\ \& fit:}                  \\ \hline
\mes\ fit                 &    4.1     \\
$\epsilon_l^u \epsilon_t^u/\epsilon_l^{sl} \epsilon_t^{sl}$ &    3.0 \\
$B$ SL branching ratios   &     4.9    \\
$D$ branching ratios      &     0.1    \\ \hline
\breco\ \& fit errors     &   \multicolumn{1}{|r@{\hspace{5mm}}|}{7.1}   \\ \hline
\multicolumn{2}{|c|}{Signal:}                \\ \hline
Composition of \Bxulnu\ decays &     3.0    \\
Hadronization             &     3.0 \\
Gluon splitting to $s{\bar{s}}$ &     2.2   \\ 
Shape function parameters & \raisebox{2.8mm}{\rule{0pt}{1ex}}$^{+1.6}_{-1.9}$\raisebox{-2mm}{\rule{0pt}{1ex}} \\
Shape function form       &     0.3    \\ \hline
Signal efficiency         &     \multicolumn{1}{|r@{\hspace{5mm}}|}{5.1}    \\ \hline
Total systematic error    &   \multicolumn{1}{|r@{\hspace{5mm}}|}{10.4}      \\ \hline
Total error               &  \multicolumn{1}{|r@{\hspace{5mm}}|}{14.9}      \\ \hline
\end{tabular}
\label{sysall}
\end{center}
\end{table}

\section{Extraction of \Vub}
\label{sec:vub}

Using the \mx-\Q\ analysis we measure the partial 
branching fraction for charmless semileptonic decays in a selected phase
space region. To translate this into a measurement of the total  
branching fraction, and therefore \Vub, we need the fraction of events
inside the measurement region 
(referred to as ``acceptance'' in the rest of the paper) 
as an external input. 

In the following we use two different theoretical calculations of 
Bauer, Ligeti and Luke~\cite{Bauer:2001yb} (BLL) and
Bosch, Lange, Neubert and Paz~\cite{Lange:2005yw,Bosch:2004th,Bosch:2004cb} (BLNP)
for calculating acceptance corrections. Both BLL and BLNP use 
operator product expansions (OPE) to calculate QCD effects.

\subsection{Results using acceptances from BLL}
\label{moments} 

Bauer, Ligeti and Luke 
perform an OPE-based calculation to second order in 
the strong coupling constant $\alpha_s$ and $b$-quark mass $m_b$. 
They focus on the region chosen for the measurement of the partial branching
fraction where non-perturbative effects are small. 
In particular, they have shown that the theoretical uncertainties of the
extrapolation to the full phase space are much reduced by restricting the
selection to regions of higher values of \Q, rather than just restricting
\mx\ to a region below the charm meson mass. 

Based on these calculations we can convert the measured $\Delta \BR(\Bxulnu)$ into \Vub\ by
\begin{equation}
\label{eq:dbrvub}
 |V_{ub}| = \sqrt{\frac{192 \pi^3}{\tau_B \, G_F^2 m_b^5}\frac{\Delta \BR(\Bxulnu)}{G}} 
\end{equation}
where $\tau_B = 1.604\pm0.012$~ps~\cite{PDG04} and $G$ is a theoretical parameter calculated in the
BLL approach~\cite{Bauer:2001yb}.  
The first factor under the square root is 
192$\pi^3/(\tau_B G_F^2 m_b^5)=0.00779$. 
To extract \Vub, we take $G=0.27$ as computed by BLL for $m_b$(1S) =
4.7~\gevcc. We then infer the $b$-quark mass in the 1S scheme from the \babar\ measurement 
of \mbk~\cite{Aubert:2004aw} by using the prescription in~\cite{Battaglia:2003in},
obtaining $m_b$(1S) = 4.74~\gevcc. G is then
recomputed by rescaling the original BLL value by the ratio
$(4.74/4.7)^9$~\cite{ligeti},
obtaining $G=0.291\pm0.055$.
The 19\% error on $G$, which turns into a 9.5\% error on \Vub, is the sum in quadrature of 
uncertainties due to: residual SF effects, higher order terms in
the $\alpha_s$ perturbative expansion, a 80~\mevcc 
uncertainty on the $b$ quark mass, and ${\cal{O}}(\Lambda^3_{\rm{QCD}}/m^3)$ terms in the OPE expansion. 
The uncertainty on the $b$ quark mass is the dominant source, contributing about 15\% to the
uncertainty on $G$. Eq.~\ref{eq:dbrvub} yields
\begin{equation}
|V_{ub}|  = (4.82 \pm 0.26_\mathrm{stat} \pm 0.25_\mathrm{syst} \pm 0.46_\mathrm{th+SF}) \times 10^{-3}.
\end{equation}

\subsection{Results using the theoretical calculations by BLNP}
\label{sec:theonew}

Bosch, Lange, Neubert, and Paz have performed calculations of the
differential decay rates for \Bxulnu\ and \Bsg.
The authors presented a systematic treatment of the SF effects,
incorporating all known corrections to the rates, and provided an
interpolation between regions of phase space that can be treated
reliably by OPE calculations and others that depend on SF. 
They have introduced a parameterization of the SF. The parameters
describing the SF cannot be calculated, rather they have to be taken
from experiment.

On the basis of these SF parameters, the partial rate for \Bxulnu\ can
be predicted for the measured phase space, and related to \Vub,
\begin{equation}
  \Delta\zeta \, \Vub^2 = 
  \int^{\mathrm{\mx^{cut}}}_{0}\int^{}_{\Q_{\mathrm{cut}}}
  \frac{d\sigma}{d\Q d\mx} d\Q d\mx,
\end{equation}                                                                                
such that
\begin{equation}
\Vub = \sqrt{\frac{\Delta{\cal{B}}(\Bxulnu)}{{\Delta \zeta \, \tau_B}}}.
\end{equation}
BLNP give results and uncertainties in terms of the reduced decay rate 
$\Delta \zeta$, defined in units of $\Vub^2$~ps$^{-1}$.
                                                                                
We rely on two measurements of these SF parameters, one based on the
photon spectrum in \Bsg\ decays, the other on moments of the hadron
mass and lepton energy spectrum in \Bxclnu\ decays.
The analysis of \Bsg\ decays can be used to determine the SF parameters in a given
renormalization scheme~\cite{Kagan:1998ym}. 
Likewise the moments of the lepton energy and hadronic invariant mass in
\Bxclnu\ decays are sensitive to the heavy quark
parameters, as shown in an OPE calculation~\cite{Gambino:2004qm} in
the kinetic scheme. 
In
both cases, the heavy quark parameters entering the calculations can be related to
\lbarsf\ and \lonesf, see {\it{e.g.}}~\cite{Neubert:2004sp,Benson:2004sg}.

The Belle Collaboration has measured the photon spectrum in \Bsg
decays~\cite{Koppenburg:2004fz} and, based on a fit to the spectrum,
has determined $\lbarsf = 0.66 \gev/c^2$ and
$\lonesf=-0.40\gev^2/c^4$~\cite{Limosani:2004jk}. They also provide
a $\Delta\chi^2=1$ contour, which we use to estimate theoretical uncertainties.

These SF parameters translate to $\mbsf = 4.52 \pm 0.07$ and 
$\mupisqsf = 0.27 \pm 0.23$~\cite{Bizjak:2005hn}.
This results in 
$\Delta\zeta = (21.6 \pm 4.0\pm^{2.4}_{2.3})\Vub^2$~ps$^{-1}$,
where the first error is due to the limited experimental knowledge of the
SF parameters and the second to theory uncertainties,
and consequently
\begin{equation}
|V_{ub}|  = (5.00 \pm 0.27_\mathrm{stat} \pm 0.26_\mathrm{syst} \pm
 0.46_\mathrm{SF} \pm 0.28_\mathrm{th}) \times 10^{-3},
\end{equation}
where the errors are due to statistics, experimental systematics, shape function parameters 
and theoretical systematics, respectively.

Alternatively, the \babar\ collaboration has determined 
\mbk\ and \mupisqk\ in the kinetic mass scheme from fits to moments
measured for \Bxclnu~\cite{Aubert:2004aw}.
The values have been translated into the SF scheme by following the
prescription in~\cite{Neubert:2004sp} resulting in
$\mbsf = 4.61 \pm 0.08$ \gevcc and $\mupisqsf = 0.15 \pm 0.07$, with a correlation
of -40\%. 
The systematic error due to the uncertainty of the SF parameters is reduced, due to the 
significantly better precision obtained in the \babar\ moments analysis. 

By using the results of the \babar\ moments analysis we get 
$\Delta \zeta = (25.04 \pm^{4.91}_{4.06\mathrm{SF}}\pm2.45_\mathrm{th})$ $\Vub^2$~ps$^{-1}$. Again, the 
error is due to the limited experimental knowledge of the
shape function parameters. 
This translates into
\begin{equation}
|V_{ub}|  = (4.65 \pm 0.24_\mathrm{stat} \pm 0.24_\mathrm{syst} \mbox{}^{+0.46}_{-0.38 \mathrm{SF}} \pm 0.23_\mathrm{th}) \times 10^{-3}.
\end{equation}

\section{Conclusions}
\label{sec:conclusions}

We have presented a study of charmless semileptonic decays and a measurement of the \Vub 
CKM matrix element, by using the combined information of the \mx-\Q\ distribution to discriminate signal and 
background and to minimize the theoretical uncertainties. 
We give a measurement of the partial branching fraction of charmless semileptonic decays $\Delta\BR(\Bxulnu)$ 
for $\mx<1.7 \gevcc$ and $\Q>8 \gevccsq$ and, by taking kinematic acceptances 
from two theoretical calculations by BLL and BLNP, extract \Vub. 

The measured partial branching fraction $\BR(\Bxulnu)$ 
in the region limited by 
$\mx < 1.7 \gevcc$, $\Q>8 \gevccsq$ is 
\begin{eqnarray}
  \begin{array}{rll}
    \lefteqn{\Delta\BR(\Bxulnu,\mx<1.7 \gevcc, \Q>8 \gevccsq)  = } \\
    & & =(0.87 \pm 0.09_{\mathrm{stat}} \pm 0.09_{\mathrm{syst}} \pm 0.01_{\mathrm{th}})\times 10^{-3}.
  \end{array}
\end{eqnarray}
We extract the CKM matrix element \Vub\ using different approaches.
With acceptances calculated using the BLL calculations, we obtain:
\begin{equation}
  |V_{ub}|^{\mathrm{BLL}}  =  (4.82 \pm 0.26_\mathrm{stat} \pm
  0.25_\mathrm{syst} 
  \pm 0.46_\mathrm{th+SF}) \times 10^{-3}. 
\end{equation}
Using the partial decay models calculated in the BLNP approach
and by taking the shape function parameters from  
the Belle photon spectrum in \Bsg and the \babar\ analysis of \Bxclnu moments,
we find:
\begin{eqnarray}
  |V_{ub}|^{\mathrm{BLNP}}_{\mathrm{Belle} \ \Bsg} & = & (5.00 \pm
  0.27_\mathrm{stat} \pm 0.26_\mathrm{syst} \pm 0.46_\mathrm{SF} \pm
  0.28_{\mathrm{th}}) \times 10^{-3}, \\  
  |V_{ub}|^{\mathrm{BLNP}}_{\mathrm{\babar}\ \bclnu}  & = & (4.65 \pm
  0.24_\mathrm{stat} \pm 0.24_\mathrm{syst} \mbox{}^{+0.46}_{-0.38 \mathrm{SF}} \pm
  0.23_{\mathrm{th}}) \times 10^{-3},
\end{eqnarray}
where the errors are due to statistics, experimental systematics, SF and theoretical systematics, respectively. 

In conclusion, the total error on \Vub is dominated by the
experimental and theoretical uncertainties of the shape function. 
Our results of \Vub using the two different
calculations of BLL and BLNP are consistent with each other.
For the BLNP calculations the two sets of shape function parameters
coming from a fit to the photon energy spectrum and to the \Bxclnu 
moments are in good agreement and thus give consistent results on \Vub. 
Results based on partial branching fraction from the lepton spectrum
and \Q\ using BLNP appear to be consistent with our measurement but
somewhat lower~\cite{Aubert:2005im}.

\section{ACKNOWLEDGMENTS}
\label{sec:Acknowledgments}
We are grateful for the 
extraordinary contributions of our \pep2\ colleagues in
achieving the excellent luminosity and machine conditions
that have made this work possible.
The success of this project also relies critically on the 
expertise and dedication of the computing organizations that 
support \babar.
The collaborating institutions wish to thank 
SLAC for its support and the kind hospitality extended to them. 
This work is supported by the
US Department of Energy
and National Science Foundation, the
Natural Sciences and Engineering Research Council (Canada),
Institute of High Energy Physics (China), the
Commissariat \`a l'Energie Atomique and
Institut National de Physique Nucl\'eaire et de Physique des Particules
(France), the
Bundesministerium f\"ur Bildung und Forschung and
Deutsche Forschungsgemeinschaft
(Germany), the
Istituto Nazionale di Fisica Nucleare (Italy),
the Foundation for Fundamental Research on Matter (The Netherlands),
the Research Council of Norway, the
Ministry of Science and Technology of the Russian Federation, and the
Particle Physics and Astronomy Research Council (United Kingdom). 
Individuals have received support from 
CONACyT (Mexico),
the A. P. Sloan Foundation, 
the Research Corporation,
and the Alexander von Humboldt Foundation.
%
Finally, we would like to thank the many theorists with whom we
have had valuable discussions, and further thank M. Neubert, B. Lange 
and G.Paz for making available for our use a computer code implementing
their calculations.

\bibliographystyle{physrev4wt}
\bibliography{note1255.bib}
\cleardoublepage

\end{document}